\documentstyle[12pt]{article}
\begin{document}
\begin{center}{\large{\bf SCHR\"ODINGER WAVE FUNCTION FOR A FREE FALLING
PARTICLE IN THE SCHWARZSCHILD BLACK HOLE}}
\end{center}
\vspace*{1.5cm}
\begin{center}
A. C. V. V. de Siqueira, I. A. Pedrosa and E. R. Bezerra de Mello
$^{*}$ \\
Departamento de F\'{\i}sica-CCEN\\
Universidade Federal da Para\'{\i}ba\\
58.059-970, J. Pessoa, PB\\
C. Postal 5.008\\
Brazil
\end{center}
\vspace*{1.5cm}
\begin{center}{\bf Abstract}
\end{center}

We use the time-dependent invariant method in a 
geometric approach (Jacobi fields) to quantize the motion of a free 
falling 
point particle in the Schwarzschild black hole. Assuming that the 
particle 
comes from infinity, we obtain the relativistic Schr\"{o}dinger wave 
function for this system.

\vspace{3cm} 

{\footnotesize \rm One of us (E. R. B. M) wishes to acknowledge 
Conselho
Nacional de
Desenvolvimento Cient\'{\i}fico e Tecnol\'{o}gico (CNPq.) for partial
financial support.\\
${}^*$ E-mail: eugenio@dfjp.ufpb.br
\newline
PACS numbers: 03.65.+; 03.65.-w, 04.90 + e, 12.25}

\newpage
 
\section{Introduction}
$         $

The study of the quantization of the motion of particles in a general 
 gravitational background space-time has attracted considerable interest 
 in the literature\cite{1}. These analyses offer the opportunity to be a 
 good theoretical laboratories in the attempt to construct an effective 
 quantum theory of gravitation.

  The General Relativity Theory assures us that every coordinate frame 
is 
 physically equivalent from a classical point of view. Thus we can 
affirm 
 that any falling particle in a gravitational field will be accelerated 
 relatively to any stationary coordinate frame defined by a Killing 
vector. 
 However, it is possible that a specific space-time has no Killing 
vector, 
 so in this case we would not have at our disposal a frame which 
enables 
 us to measure this acceleration. In such a situation, the best we 
can do
 is to take two 
 different particles and measure their relative acceleration. This 
relative 
 accelaration will be given by the equation which governs the Jacobi 
 fields\cite{2}. These equations, which are classical, can indicates 
us 
 an analogous quantum mechanical system which permits us to study the 
corresponding quantum motion of the particle in this general 
gravitational 
background space-time. This is what happers for a free falling 
particle in 
the Schwarzschild space-time.
As we shall see, the classical Jacobi field equations for a free 
falling 
particle in the Schwarzschild metric space-time is similar to a classical 
harmonic oscilator with a time-dependent frequence. Fortunately there
exists
a well 
defined procedure to quantize systems like that through the use of 
invariants\cite{3,4}. This technique has been demonstrated to be a 
powerful 
one to quantize such systems, and to yeld a closed expression for the 
Schr\"odinger wave function.

In the present paper we shall use the theory of explicitly 
time-dependent invariants to quantize the free motion of a point 
particle 
in the Schwarzschild black hole using Jacobi fields. The time 
parameter in 
the relativistic Schr\"{o}\-dinger equation will be the affine 
parameter,
the
proper time of the particle in this manifold. As will become clear, 
this geometric approach has two advantages: it avoids the ordering 
problem 
of differential operators in a general background 
space-time\cite{5}, and it also avoids the horizon problem in our 
specific 
space-time, because, as we shall see, our relativistic wave function 
is
well
defined at $r=2M.$

So the main objective of this paper is to study from quantum point 
of 
view the motion of a free falling particle in the Schwarzschild black 
hole.

This paper is organized as follows. In Sec. $2$ we introduce the Jacobi 
field equations for general case. In Sec. $3$ the invariant method and 
the relativistic Schr\"{o}dinger equation associated with this 
geometric 
approach are analized. In Sec. $4$ we apply the formalism reviewed 
in the 
Sec. $3$ to study the motion of a free point particle falling in the 
Schwarzschild black hole. In Sec. $5$ we summarize the main results of 
this work.

\renewcommand{\theequation}{\thesection.\arabic{equation}}
\section{\bf Jacobi Fields}
\setcounter{equation}{0}
$         $

In this section we briefly review the Jacobi fields and their 
respective 
differential equation for a general manifold\cite{6}.

Let us consider a differentiable manifold, $\cal{M}$, and two 
structures 
defined on $\cal{M}$, namely affine connection, $\nabla$ , and 
Riemann 
tensor, $R$ , related by the generic equation

\begin{equation}
\label{21}
R(X,Y)Z=\nabla _{X}\nabla _{Y}Z-\nabla _Y\nabla _{X}Z-\nabla _{[X,Y]}Z,
\end{equation}
where $X,Y$ and $Z$ are vector fields in the tangent space.

The torsion tensor can be defined by

\begin{equation}
\label{22}
T(X,Y)=\nabla _{X}Y-\nabla _{Y}X-[X,Y], 
\end{equation}
where $\nabla _{X}Y$ is the covariant derivative of the field $Y$ 
along the 
$X$ direction. We also can define the Lie derivative by

\begin{equation}
\label{23}
{\cal{L}}_{X}Y=[X,Y]. 
\end{equation}
For a connection free of torsion we can write

\begin{equation}
\label{24}
{\cal{L}}_{X}Y=\nabla _{X}Y-\nabla _{Y}X.
\end{equation}
On the other hand, we are interested in constructing a family of 
curves 
$\lambda (t,s)$, moving each point $\lambda (t)$ a distance $s$ along 
the 
integrals curves of $V=(\frac \partial {\partial t})_{\lambda (t,s)}$. 
Let 
the vector field $Z$ be equal to $(\frac \partial {\partial t})_
{\lambda (t,s)},$ so that we have a family of curves with the condition 

\begin{equation}
\label{25}
{\cal{L}}_{V}Z=0. 
\end{equation}

From (\ref{25}), we can easily see that the equations which govern the 
Jacobi field can be written in the form

\begin{equation}
\label{26}
\nabla _V\nabla _VZ+R(...,V,Z,V)-\nabla _Z\nabla _VV=0.  
\end{equation}

In this paper we are interested in studying the quantum 
motion of a point particle interacting with a background gravitational 
field, so, the particle's world lines are geodesics in this space-time, 
whose equations are given by

\begin{equation}
\label{27}
\nabla _{V}V=0.  
\end{equation}

In this case the Jacobi equations are reduced to a geodesic 
desviation, assuming a simpler form, and the Fermi derivative 
$\frac{D_FZ}{\partial s}$ coincides with the usual covariant 
derivative\cite{7}:

$$
\frac{D_FZ}{\partial s}=\frac{DZ}{\partial s}=\nabla _VZ. 
$$
Because the particle possesses a non vanishing rest mass, it is 
convenient 
to define the tangent vector $V$ as a time-like one, so that 
$g(V,V)=-1.$ 
Considering $e_o,$ $e_1,$ $e_2,$ and $e_3$ be an orthogonal basis 
of the 
space-time, in the rest frame of the particle, $V=e_o.$ So, taking 
the 
Fermi-Walker transport into Eq. (\ref{26}), we can write

\begin{equation}
\label{28}
\frac{d^2Z_i}{d\tau ^2}+R_{0i0j}Z_j=0, 
\end{equation}
where $\tau $ is, in general, an affine parameter, which in our case 
is the
proper time of the particle and $Z_i$ the component of the space-like 
vector $Z$, with $g(Z,V)=0$. We can observe that the Fermi-Walker 
transport 
for our case, maps the stationary  Schwarzschild coordinate system 
into 
another one, contructed by successive Lorentz boosts, where the 
particle, 
whose motion
is governed by (\ref{28}), interacts with the curvature of the 
space-time 
as an external
force. The system constructed by the successive Lorentz frames is not 
conservative, so at quantum level, the probability density is 
explicitly 
time-dependent
as we one can see from the wave function given in Sec.$4$.

For the case where $R_{0i0j}$ is diagonal in the spatial indices
\cite{8}, 
the Eq. (\ref{28}) becomes

\begin{equation}
\label{29}
\frac{d^2Z_i}{d\tau ^2}+R_{0i0i}Z_i=0, 
\end{equation}
with no summation indices assumed.

As we have already mentioned, in this paper we shall specialize to the
Schwarzschild black-hole case. A more general analysis is now under 
investigation.

\renewcommand{\theequation}{\thesection.\arabic{equation}}
\section{\bf Time-Dependent Invariants and the Relativistic Schr\"odinger 
Equation}

\setcounter{equation}{0}
$         $

In this section we shall briefly review some important points related
 with 
the invariant method which we shall employ to quantize the motion of 
a free 
falling particle in a specific space-time through a geometric 
approach.

Let as now consider the following time-dependent Hamiltonian

\begin{equation}
\label{31}
H(\tau )=\sum_{j=1}^3\left[ \frac{P_j^2}{2m}+\frac m{2
}R_{0i0i}(\tau
)Z_j^2\right] ,  
\end{equation}
where $P_{i}$ is the canonical momentum conjugated to the coodinate
$Z_{i}$.
We can notice that $R_{0i0i}$ is not a variable on phase space, 
and that 
the equation above is formally similar to an anisotropic 
time-dependent 
harmonic oscilator, whose classical equation for any direction 
is given 
by Eq.$(2.9)$.

In order to quantize this system we shall employ the invariant method 
which depends on the so called invariant operator. Fo our system this 
operator is expressed by \cite{4,9,10}

\begin{equation}
\label{32}
I(\tau )=\frac 1{2}\sum_{i=1}^3\left[ K_i\left( \frac{Z_i}{\sigma _i}
\right)
^2+\left( \sigma _{i}P_i-m\dot{\sigma}_i^{}Z_i\right) ^2\right] , 
\end{equation}
where $K_{i}$ is an arbitrary constant and $\sigma_{i}(\tau )$ a 
c-number 
obeying the inhomogeneous auxiliary equation

\begin{equation}
\label{33}
\frac{d^2\sigma _i}{d\tau^{2}}+R_{0i0i}(\tau )\sigma_i=\frac{K_i}{m^2
\sigma _
i^{3}}.
\end{equation}

The invariant operator $I(\tau )$ satisfies the equation \cite{3,4}

\begin{equation}
\label{34}
\frac{dI(\tau )}{d\tau }=\frac{\partial I(\tau )}{\partial \tau }+\frac
1{i\hbar }\left[I,H\right] =0.  
\end{equation}

Because the operators  $I(\tau )$ and  $H(\tau )$ are written as the 
sum of 
three independent expressions in each direction, the solution of the 
equation of motion can be written as a product of solutions, each 
refering
 to one direction. So we shall focus our attention just to one 
direction. 
Let $\phi_{\lambda_{i}}(Z_{i},\tau )$ be the eigenfunction of the 
operator  $I_{i}(\tau )$ with eigenvalue $\lambda_{i}$. Thus we have:

\begin{equation}
\label{35}
I_i(\tau )\phi _{\lambda _i}(Z_i,\tau )=\lambda _i\phi _{\lambda _i}.
\end{equation}
Next let us consider the Schr\"odinger equation

\begin{equation}
\label{36}
i\hbar \frac \partial {\partial \tau }\Psi _{\lambda _i}(Z_i,\tau )=H_i
(\tau)\Psi _{\lambda _i}(Z_i,\tau ), 
\end{equation}
with

\begin{equation}
\label{37}
H_i(\tau )=-\frac{\hbar ^2}{2m}\frac{\partial ^2}{\partial Z_i^2}+\frac
1{2}mR_{0i0i}(\tau )Z_i^{2}, 
\end{equation}
where we have used the operator $P_i$ in its differential form.

In order to obtain the solution for the relativistic  Schr\"odinger 
equation\footnote{Although Eq.$(3.6)$ is not in a covariant form, it is 
in fact derived from the relativistic time-dependent Hamiltonian 
Eq.$(3.1)$} $(3.6)$, we shall use the Risenfeld an Lewis (RL) method 
\cite{3,4}, which permits to express $\Psi_{\lambda_{i}}$ in terms of 
another function $\phi_{\lambda_{i}}$ by

\begin{equation}
\label{38}
\Psi _{\lambda _i}(Z_i,\tau )=\exp [i\alpha _{\lambda _i}(\tau )]\phi_
{\lambda _i}(Z_i,\tau ),  
\end{equation}
$\alpha_{\lambda_{i}}(\tau )$ being the phase factor which obeys the 
equation

\begin{equation}
\label{39}
<\phi _{\lambda _i}\mid \phi _{\lambda _j}>\hbar \frac{\partial \alpha
_{\lambda _j}}{\partial \tau }=<\phi _{\lambda _i}\mid i\hbar \frac 
\partial{\partial \tau }-H(\tau )\mid \phi _{\lambda _j}>. 
\end{equation}

The new function $\phi_{\lambda_{i}}$, on the other hand, is obtained, 
by 
the unitary transformation,

\begin{equation}
\label{310}
\phi _{\lambda _i}^{\prime }=U_i\phi _{\lambda _i}=\exp 
\left[ -\frac{im}
{2\hbar }\frac{\dot{\sigma}}{\sigma}Z_i\right]\phi _{\lambda _i}
\end{equation}
with the solution of the eigen-values equation

\begin{equation}
\label{311}
I_i^{\prime }\phi _{\lambda _i}^{\prime }=\lambda _i\phi _{\lambda
_i}^{\prime }
\end{equation}
where $I_{i}^{'}$ is operator

\begin{equation}
\label{312}
I_i^{\prime }=-\frac{\hbar ^2}2\sigma _i^2\frac{\partial ^2}{\partial 
Z_i^2}+\frac{ K_i}2\frac{Z_i^2}{\sigma _i^2}\qquad ,  
\end{equation}
similar to the harmonic oscilator Hamiltonian.

Again, defining a new variable $X_{i}=\frac{Z_{i}}{\sigma_{i}}$ and 
the
field
$\phi_{\lambda_{i}}=\sigma_{i}^{1/2}\phi_{\lambda_{i}}^{'}$, we get 
the 
eigenvalue equation

\begin{equation}
\label{313}
\left( -\frac{\hbar ^2}2\frac{d^2}{dX_i^2}-\frac{K_i}
{2}X_i^2\right) \varphi _{\lambda _i}=\lambda _i\varphi _{\lambda _i}.
\end{equation}

Thus the objective of the invariant method (RL) is to express a 
general 
solution of a time-dependent Schr\"odinger equation in terms of some 
well 
known eigenvalue differential equation. For our specific case, this 
was an 
harmonic oscilator differential equation. The only problem now is to 
solve 
the inhomogeneous auxiliary equation $(3.4)$. Fortunatly, although 
this 
equation is not linear, we were able to  obtain, for the problem that 
we 
shall analyse in the next section, different solutions 
for $\sigma_{i}(\tau )$ for positive and negative values of the 
constant $K_{i}$, which, on the other hand, leads us to an attractive 
and repulsive, respectively, harmonic potential into $(3.13)$.

\renewcommand{\theequation}{\thesection.\arabic{equation}}
\section{\bf The Relativistic Schr\"odinger Wave Function for a 
 Particle in the Schwarzschild Black Hole}
\setcounter{equation}{0}
$         $

The exterior line element of the Schwarzschild black-hole is given by

\begin{equation}
\label{41}
ds^2=-\left( 1-\frac{2M}r\right) dt^2+\left( 1-\frac{2M}r\right)
^{-1}dr^2+r^2d\Omega ^2.  
\end{equation}

In accordance with Eq.$(2.9)$, the non-vanishing Riemann tensor 
components 
of interest in this space-time are:

\begin{equation}
\label{42}
R_{0101}=-\frac{2M}{r^3}\qquad , 
\end{equation}
and

\begin{equation}
\label{43}
R_{0202}=R_{0303}=\frac M{r^3}. 
\end{equation}
where the above relations are invariant under a boost along the radial 
direction for r$>2$M, and still remain valid for r$\leq2$M \cite{11}.

As we have pointed out at the end of the last section, in order to 
obtain 
the Schr\"odinger wave function  $\Psi_{\lambda_{i}}(Z_{i},\tau )$, 
we have 
first to solve the auxiliary equation $(3.4)$, which, on the other 
hands, 
depends on the function $R_{0i0i}(\tau )$. This function can be 
constructed 
through the differential equations obeyed by the classical Jacobi 
fields, 
which for this geometry read

\begin{eqnarray}
\label{44}
\frac{d^{2}Z_i}{d\tau ^2}-\frac{2M}{r^3}Z_i=0, 
\end{eqnarray}
and

\begin{equation}
\label{45}
\frac{d^{2}Z_j}{d\tau ^2}+\frac M{r^3}Z_j=0,  
\end{equation}
where the index j assumes values $2$ e $3$ in the last equation. Now 
assuming that the test particle is at rest at infinity, it is possible 
to 
show \cite{12} the relationship between the radial coordinate and the 
proper time

\begin{equation}
\label{46}
\frac{dr}{d\tau }=-\sqrt{\frac{2M}r} 
\end{equation}
whose solutions is

\begin{equation}
\label{47}
\tau =-\frac 2{3\sqrt{2M}}r^{3/2}+const.  
\end{equation}

The constant in the equation above can be absorbed in the redefinition of 
the zero of the proper time. We can also, see from this equation that 
for r$\rightarrow-\infty$, the propers time $\tau\rightarrow -\infty$.

Now we can rewrite Eq's $(4.2)$ and $(4.3)$ in terms of the affine 
parameter, the proper time of the particle.

\begin{equation}
\label{48}
R_{0101}=-\frac{4}{9\tau^{2}}\qquad, 
\end{equation}
and

\begin{equation}
\label{49}
R_{0202}=R_{0303}=\frac{2}{9\tau^{2}}\qquad , 
\end{equation}

The classical solutions for Eqs.$(4.4)$ and $(4.5)$, in terms of the 
affine parameter $\tau$ are given by 

\begin{equation}
Z_{1}(\tau)=c_{i} \tau^{4/3} \nonumber 
\end{equation} 
and

\begin{equation}
Z_{j}(\tau)=c_{j} \tau^{2/3}, \nonumber 
\end{equation}
where $c_{i}$ and $c_{j}$, for $j=2$ and $3$, are constants. One can 
see that 
the expressions above are well defined at $r\leq 2M$. 

After this brief analysis, let us return to the construction of the 
Schr\"odinger wave function by means of the invariant method. In order 
to 
do that we need to solve first the auxiliary equations

\begin{equation}
\label{410}
\frac{d^2\sigma _1}{d\tau ^2}-\frac 4{9\tau ^2}\sigma _1=\frac{K_1}{m^2
\sigma _i^3} 
\end{equation}
and

\begin{equation}
\label{411}
\frac{d^2\sigma _j}{d\tau ^2}+\frac 2{9\tau ^2}\sigma _j=\frac{K_j}{m^2
\sigma _i^3}\qquad ,
\end{equation}
where $j=2$ and $3$.

For the case $K_{i}>0$, we obtained the following solutions:

\begin{equation}
\label{412}
\sigma_{1}(\tau )=\left[(1+a_{1}^{2})c_{1}^{2}\mid\tau\mid^{8/3}+
\frac{6a_{1}K_{1}^{1/2}\mid\tau\mid}{5}+\frac{9K_{1}}{25c_{1}^{2}\mid\tau
\mid^{2/3}}\right]^{1/2},
\end{equation}
and

\begin{equation}
\label{413}
\sigma_{j}(\tau )=\left[(1+a_{j}^{2})c_{j}^{2}\mid\tau\mid^{4/3}+6a_{j}K
_{j}^{1/2}\mid\tau\mid+9K_{j}\mid\tau\mid^{2/3}\right]^{1/2},
\end{equation}
where $a_{1}$ and $a_{j}$ are constants of integration.

For the case $K_{i}\leq 0$, we obtained

\begin{equation}
\label{414}
\sigma_{1}(\tau )=i\left[\mid b_{1}\mid c_{1}\mid\tau\mid^{8/3}+\frac{6
\mid K_{1}\mid^{1/2}}{5}\mid\tau\mid\right]^{1/2},
\end{equation}
and

\begin{equation}
\label{415}
\sigma_{j}(\tau )=i\left[\mid\ b_{j}\mid c_{j}\mid\tau\mid^{4/3}+6\mid K
_{j}\mid^{1/2}\mid\tau\mid\right]^{1/2},
\end{equation}
again $b_{1}$ and $b_{j}$ are constants of integration.

Our next step in the construction of the wave function is to obtain the 
phase factor and the unitary transformation, and then to solve the 
eigenvalue equation $(3.13)$. For the first two factors the knowledge 
of $\sigma_{i}(\tau )$ is essential. The phase factor is given by

\begin{equation}
\label{416}
\alpha_{\lambda_{i}}(\tau )=\alpha_{\lambda_{i}}^{(0)}\int\frac{d\tau}
{\sigma_{i}^{2}}(\tau ),
\end{equation}
 $\alpha_{\lambda_{i}}^{(0)}$ being a constant, and the unitary operator 
is 
given by Eq.$(3.10)$.

Finally in order to solve $(3.13)$ we have to consider the two 
cases $K_{i}>0$ and $K_{i}\leq0$. For the first case, Eq. $(3.13)$ 
becomes 
equivalent to an harmonic oscilator, with a discrete set of 
eigenvalues $\lambda_{n}$. For negative and vanishing values of $K_{i}$, 
Eq. $(3.13)$ becomes equivalent to an inverted harmonic oscilator and a 
free system, respectively. Let us first analyse the discrete case:

i) For $K_{i}>0$, the solutions for $\sigma_{i}(\tau )$ are given by 
Eq. $(4.12)$ and $(4.13)$, and the wave functions are

\begin{eqnarray}
\Psi _{\lambda _i}(Z_i,\tau )=\frac{\exp^{i\alpha_{\lambda_{i}}(\tau )}}
{\left[\pi^{1/2}\hbar^{1/2}n_{i}!2^{n/2}\right]}
\frac{exp\left[-\frac{m}{2i\hbar}\frac{\dot{\sigma_{i}}}{\sigma_{i}}Z_{i}
^{2}\right]}{\sigma_{i}^{1/2}}exp\left[-\frac{Z_{i}^{2}}{2\hbar\sigma_{i}
^{2}}\right].\nonumber
\end{eqnarray}

\begin{equation}
\label{417}
H_{\lambda_{i}}\left[\frac{Z_{i}}{\hbar^{1/2}\sigma_{i}}\right],
\end{equation}
where the phase  $\alpha_{\lambda_{i}}(\tau )$ has been given above and
$H_{\lambda_{i}}$ are the Hermite polynomials.

ii) For $K_{i}<0$ we also have the solutions for  $\sigma_{\lambda_{i}}
(\tau )$ given by Eq's $(4.14)$ and $(4.15)$, and the respective wave
 function are

\begin{equation}
\label{418}
\Psi _{\lambda _i}(Z_i,\tau )=\exp^{i\alpha_{\lambda_{i}}(\tau )}\frac{exp
\left[-\frac{m}{2i\hbar}\frac{\dot{\sigma_{i}}}{\sigma_{i}}Z_{i}^{2}
\right]}{\sigma_{i}^{1/2}}D_{\nu_{i}}\left[\frac{(1+i)Z_{i}}{\hbar^{1/2}
\sigma_{i}}\right] ,
\end{equation}
where $2\nu_{i}=-1-i\lambda_{i}/\hbar$, $\lambda_{i}$ being a continuos 
parameter, and $D_{\nu}(Y)$ are the parabolic cylinder functions 
\cite{13}.

iii) Finally for $K_{i}=0$ the soluctions for $\sigma_{i}(\tau )$ are:

\begin{equation}
\label{419}
\sigma_{1}(\tau )=c_{1}\mid\tau\mid^{4/3},
\end{equation}
and

\begin{equation}
\label{420}
\sigma_{2}(\tau )=c_{j}\mid\tau\mid^{2/3}.
\end{equation}

The wave function are now given by:

\begin{equation}
\label{421}
\Psi_{\lambda_{i}}(Z_{i},\tau )=\frac{exp\left[i(\frac{\dot{\sigma}}
{2\hbar\sigma_{i}}Z_{i}^{2}+\alpha_{\lambda{i}}(\tau )+\frac{\lambda
_{i}Z
_{i}}{\sigma_{i}})\right]}{\sigma_{i}^{1/2}}
\end{equation}

So the complete Schr\"odinger wave function which describes the radial 
motion of a test particle on the Schwarzschild space-time, is a linear 
combination of
 the solutions $(4.19)$, $(4.20)$ and $(4.23)$.

\section{Concluding Remarks}

In this paper we have presented a new approach, a geometric one, to
analyse,
under a quantum point of view, the motion of a point particle 
interacting 
with an external background gravitational field. As a direct application 
of 
this formalism, we have investigated the motion of a point particle 
in the 
Schwarzschild black hole, and for this case we have obtained the 
relativistic Schr\"odinger wave function, using the invariant method. 
We 
would like to emphasize that, because our system is not conservative, 
there is no defined energy for the particle, and the probability 
density 
depends explicitly on the proper time $\tau$.

This geometric approach can also be applied to other kinds of manifold, 
actually under investigation by us, and that some of its advantages are 
very clear like, for example, (i) it avoids the ordering problem for 
differential operators which generally appears in the Schr\"odinger 
equation defined on a curved space-time \cite{5}, and (ii) it also 
avoids 
the horizon problem at r$=2$M for the Schwarzschild space-time because 
our 
wave function is well defined on this region.


\end{document}